\documentclass[preprint]{aastex}

\shorttitle{Spatial distribution of $\mathrm{H_2O}$ and $\mathrm{CO_2}$ ices in M~82}
\shortauthors{M. Yamagishi et al.}

\begin{document}

\title{Difference in the spatial distribution between $\mathrm{H_2O}$ and $\mathrm{CO_2}$ ices in M~82 found with $AKARI$}

\author{Mitsuyoshi Yamagishi\altaffilmark{1}, Hidehiro Kaneda\altaffilmark{1}, Daisuke Ishihara\altaffilmark{1}, Shinki Oyabu\altaffilmark{1}, Takashi Onaka\altaffilmark{2}, Takashi Shimonishi\altaffilmark{3}, Toyoaki Suzuki\altaffilmark{4}, and Young Chol Minh\altaffilmark{5}}

\email{yamagishi@u.phys.nagoya-u.ac.jp}

\altaffiltext{1}{Graduate School of Science, Nagoya University, Furo-cho, Chikusa-ku, Nagoya 464-8602, Japan}
\altaffiltext{2}{Graduate School of Science, The University of Tokyo, 7-3-1 Hongo, Bunkyo-ku, Tokyo 113-0033, Japan}
\altaffiltext{3}{Graduate School of Science, Kobe University, 1-1 Rokkodai, Nada-ku, Kobe 657-8501, Japan}
\altaffiltext{4}{Institute of Space and Astronautical Science, Japan Aerospace Exploration Agency, Chuo-ku, Sagamihara 252-5210, Japan}
\altaffiltext{5}{Korea Astronomy and Space Science Institute, 776 Daedeok-daero, Yuseong, Daejeon 305-348, Korea}

\begin{abstract}
With $AKARI$, we obtain the spatially-resolved near-infrared (2.5 -- 5.0 $\micron$) spectra for the nearby starburst galaxy M~82.
These spectra clearly show the absorption features due to interstellar ices.
Based on the spectra, we created the column density maps of $\mathrm{H_2O}$ and $\mathrm{CO_2}$ ices.
As a result, we find that the spatial distribution of $\mathrm{H_2O}$ ice is significantly different from that of $\mathrm{CO_2}$ ice; $\mathrm{H_2O}$ ice is widely distributed, while $\mathrm{CO_2}$ ice is concentrated near the galactic center.
Our result for the first time reveals variations in $\mathrm{CO_2/H_2O}$ ice abundance ratio on a galactic scale, suggesting that the ice-forming interstellar environment changes within a galaxy.
We discuss the cause of the spatial variations in the ice abundance ratio, utilizing spectral information on the hydrogen recombination Br$\alpha$ and Br$\beta$ lines and the polycyclic aromatic hydrocarbon 3.3 $\micron$ emission appearing in the $AKARI$ near-infrared spectra.

\end{abstract}



\section{Introduction}

Absorption features due to interstellar ices are observed in near- and mid-infrared spectra of the interstellar medium (ISM).
Generally, ices are formed on the surface of dust in dense molecular clouds ($T$ $\sim$ 10~K, $n_{\mathrm{H}}\sim10^4~\mathrm{cm^{-3}}$), forming ice mantles.
In a near-infrared (NIR) spectrum, we can observe deep absorption features particularly due to $\mathrm{H_2O}$ ice at 3.05 $\micron$ and $\mathrm{CO_2}$ ice at 4.27 $\micron$.
These interstellar ices have many pieces of information on the interstellar environment.
Among various ices, $\mathrm{CO_2}$ ice is one of the most important ones as a probe of the interstellar environment.
That is because $\mathrm{CO_2}$ ice is a secondary product unlike $\mathrm{H_2O}$ and $\mathrm{CO}$ ices which are primarily formed on dust grains.
A formation process of $\mathrm{CO_2}$ ice indicated by the recent model and experimental analyses (Garrod \& Pauly 2011; Noble et al. 2011) is as follows:
\begin{equation}
\mathrm{CO} + \mathrm{OH} \rightarrow \mathrm{CO_2} + \mathrm{H}.
\end{equation}
Related to the formation process, both energetic and non-energetic ones are proposed.
As an energetic process, Watanabe \& Kouchi (2002) showed that $\mathrm{CO_2}$ ice is produced from a $\mathrm{H_2O}$-CO ice mixture by UV radiation, based on the laboratory experiments.
Alternatively, Oba et al. (2010) experimentally demonstrated a non-energetic process, in which the abundance of $\mathrm{CO_2}$ ice varies with dust temperature since the process is a grain surface reaction and a high dust temperature changes the mobility of the OH radical. 
Thus, these processes suggest that the abundance of $\mathrm{CO_2}$ ice may have information on the UV radiation environment or the dust temperature.
Additionally, ices have different sublimation temperatures (e.g. 90 K and 50 K for $\mathrm{H_2O}$ and $\mathrm{CO_2}$, respectively; Tielens 2005).
Therefore we can estimate the thermal history of dust from the presence or absence of ices.

Hence, interstellar $\mathrm{CO_2}$ ice is a useful tracer of the interstellar environment.
However, studies of $\mathrm{CO_2}$ ice, especially in nearby galaxies, have not yet been intensively performed because $\mathrm{CO_2}$ ice is not observable with ground-based telescopes due to the atmospheric absorption.
The detections of $\mathrm{CO_2}$ ice in our Galaxy are reported for some young stellar objects (Gibb et al. 2004) and several regions of the diffuse ISM (e.g, Taurus: Bergin et al. 2005, Ophiuchus: Teixeira \& Emerson 1999, and IC~5146: Chiar et al. 2011).
With {\it Infrared Space Observatory}, Spoon et al. (2000) detected the absorption features due to $\mathrm{H_2O}$, $\mathrm{CO_2}$, CO, and XCN ices in the nearby galaxy NGC~4945.
These previous observations were performed only for discrete spots.
In order to investigate $\mathrm{CO_2}$ ice more effectively, it is valuable to perform mapping observations of ices on a large scale.

Spectroscopy with the $AKARI$/{\it Infrared Camera} ($IRC$; Murakami et al. 2007; Ohyama et al. 2007) enables us to efficiently study ices in the NIR (2.5--5.0 $\micron$), which is unique in terms of continuously covering the wavelength range with high sensitivity.
Yamagishi et al. (2011) showed the NIR spectra with $\mathrm{H_2O}$ and $\mathrm{CO_2}$ ices in the central $\sim$600 pc region of NGC~253; the spatial distribution of $\mathrm{H_2O}$ ice is found to be similar to that of $\mathrm{CO_2}$ ice.
In this paper, we present the result for M~82.
Using the NIR spectra of M~82, Yamagishi et al. (2012) already discussed variations of the polycyclic aromatic hydrocarbon (PAH) 3.3 $\micron$ feature and the aliphatic sub-features at 3.4--3.6 $\micron$ in the galactic  disk and halo regions.
Here we focus on $\mathrm{H_2O}$ and $\mathrm{CO_2}$ ices to investigate their spatial distribution in the central $\sim$1 kpc region.
Unlike the case of NGC~253, we clearly reveal difference in the spatial distribution between $\mathrm{H_2O}$ and $\mathrm{CO_2}$ ices, which gives physical implications for the ice-forming interstellar environment in M~82.


\section{Observations and Data Reduction}

In all the observations, we used a grism spectroscopic mode (Ohyama et al. 2007) to obtain 2.5--5.0 $\micron$ spectra.
A summary of the observations and sub-slit apertures used in the present study is listed in Table~1, where regions A, B, and C are those defined in Yamagishi et al. (2012), covering the central $\sim$1 kpc region at a distance of M~82 (3.53 Mpc; Karachentsev et al. 2002).
The basic reduction processes are the same as those in Yamagishi et al. (2012).
Depending on the signal-to-noise (S/N) ratios, we divided the slit aperture area into sub-apertures as listed in Table~1, and derived a spectrum from each sub-aperture.
We shifted the slit sub-aperture position by 1 pixel (1$\farcs$46) one after the other to consecutively derive spectra along the slit.
Two independent observations were performed for region A, and the two spectra are combined to improve the S/N ratios.
Finally, smoothing with a boxcar kernel of $\sim0.03$ $\micron$ was applied to each spectrum.
As a result, we extract 77 spectra in total.
We consider errors given by the pipeline plus wavelength-dependent background noise; the latter errors are evaluated as 1$\sigma$ standard deviations every 0.2 $\micron$ wavelength range for 82 spectra of the blank sky.

\section{Result}

Figure~1 shows examples of the spectra extracted from the center of M~82, which reveal variations of the spectra in the direction to the south east from the galactic center.
The intensities of the spectra monotonically decrease with the distance from the galactic center.
Spectral features due to $\mathrm{H_2O}$ ice, $\mathrm{CO_2}$ ice, the PAH 3.3 $\micron$ feature, aliphatic sub-features at 3.4-3.6 $\micron$, Br$\alpha$  at 4.05 $\micron$, Br$\beta$ at 2.63$\micron$, and Pf$\beta$ at 4.65 $\micron$ are clearly detected.
As for the ices, $\mathrm{H_2O}$ ice is significantly detected in all the observed regions.
In contrast, $\mathrm{CO_2}$ ice is not detected at regions distant from the galactic center, where $\mathrm{H_2O}$ ice is still clearly detected.
In order to quantify the spatial distribution of the ices, we evaluate the column densities of the ices and create the column density maps using all the extracted spectra.


Figures~2a and 2b show the resultant column density maps of $\mathrm{H_2O}$ and $\mathrm{CO_2}$ ices.
In order to evaluate absorption features due to $\mathrm{H_2O}$ ice, we fit spectral continuum and emission/absorption features simultaneously using the wavelength ranges of 2.55--4.20, 4.35--4.40, and 4.90--4.95 $\micron$.
We adopted Lorentzian functions to fit the PAH 3.3 $\micron$ feature and aliphatic sub-feature at 3.4 $\micron$, and Gaussian functions to fit Br$\alpha$, Br$\beta$, Pf$\gamma$, and aliphatic sub-feature at 3.5 $\micron$.
The former and latter functions were used for resolved and unresolved spectral features, respectively.
As a continuum component, we used a multi-temperature (200, 400, 800, 1600, and 3200 K) blackbody model which represents warm dust emission and photospheric emission from cool stars.
The absorption feature due to $\mathrm{H_2O}$ ice is represented by a Gaussian function.
In evaluating $\mathrm{CO_2}$ ice, we used only a wavelength range of 4.12--4.42 $\micron$ to estimate local continuum levels for the relatively narrow feature as accurately as possible.
We used simple linear and Gaussian functions as continuum and absorption, respectively.
The column density, $N$, of each ice is then derived from the equation
\begin{equation}
N = \int \tau d\nu/A,
\end{equation}
where $A$, $\tau$, and $\nu$ are the band strength of each ice feature measured in a laboratory, an optical depth derived from the fitting, and a wavenumber ($\mathrm{cm^{-1}}$), respectively.
The band strengths of $2.0 \times 10^{-16}$ $\mathrm{cm~molecule^{-1}}$ and $7.6 \times 10^{-17}$ $\mathrm{cm~molecule^{-1}}$ are used for $\mathrm{H_2O}$ and $\mathrm{CO_2}$ ices, respectively (Gerakines et al. 1995).
In this calculation, we assume that the continuum and emission features are in the background and absorbed by the ices.
For the regions with significant detection of the ices ($>3\sigma$), we created the spectral maps of the column densities of the ices.


Figures~2a and 2b indicate that the ices are abundant near the galactic center.
However the spatial distribution of the ices is clearly different from that of PAHs (Kaneda et al. 2010), although both are thought to represent the neutral ISM.
The spatial distribution of PAHs has a very sharp peak in the galactic center, and it is close to symmetry with respect to the galactic center and disk.
Whereas the spatial distribution of the ices is biased toward the northeast from the center.
Such biased distribution of the ISM in the center of M~82 was also reported in the $^{12}\mathrm{CO}$ ($J$=1--0) emission (Walter et al. 2002).
In Fig.~2c, we compare the spatial distribution of the total ($\mathrm{H_2O}$+$\mathrm{CO_2}$) ice with that of CO.
The figure shows that the spatial distribution of the total ice is globally correlated with that of CO and the peak position of the ice corresponds to the bright ridge seen in the CO map.
Thus, it is clear that the spatial distribution of ices is similar to that of molecular clouds rather than that of PAHs, reflecting that ices are mainly present in dense molecular clouds which are observed in the CO emission, but not in photo-dissociation regions (PDRs) which are observed in the PAH emission.

In Figs.~2a and 2b, we also find a clear difference in the spatial distribution of the ices; $\mathrm{H_2O}$ ice is significantly detected in all the observed regions, while $\mathrm{CO_2}$ ice is detected only in smaller areas near the galactic center.
These trends are consistent with the changes of the spectra seen in Fig.~1.
In the previous study of the ices in NGC~253, there is no clear difference in the spatial distribution between $\mathrm{H_2O}$ and $\mathrm{CO_2}$ ices (Yamagishi et al. 2011).
Our result for the first time reveals variations in $\mathrm{CO_2/H_2O}$ ice abundance ratio on a galactic scale, suggesting that the ice-forming interstellar environment changes within a galaxy.

\section{Discussion}

We quantitatively examine the relation between the abundance of the ices and the properties of the ISM using the spectral features in the NIR.
Figure~3a shows the column densities of the total ice plotted against the PAH 3.3 $\micron$ feature intensity.
The figure shows a rather loose correlation ($R=+0.61$ for independent 14 regions with significant detection of $\mathrm{H_2O}$ and $\mathrm{CO_2}$ ices) as can also be expected from Fig.~2.
Thus it reconfirms that the ices are not mainly present in PDRs.
We also evaluate the amount of the ISM via the interstellar extinction.
Figure~3b shows the abundance of the total ice plotted against Br$\beta$/Br$\alpha$ ratios and $A_{\mathrm{V}}$.
In the conversion from Br$\beta$/Br$\alpha$ ratios to $A_{\mathrm{V}}$, we assume an extinction law in the NIR of $A_{\lambda} \propto \lambda^{-1.85}$ (Landini et al. 1984), total-to-selective extinction ratio, $R_\mathrm{V}$, of 3.1, and the intrinsic line ratio in Case B, temperature of $3\times10^{4}$~K, and electron density of $10^4$ $\mathrm{cm^{-3}}$ (Storey \& Hummer 1995).
By using background stars for a dark cloud in our Galaxy, Whittet et al. (1988; 2007) reported a tight correlation between $A_{\mathrm{V}}$ and column densities of the ices, which is also plotted in Fig.~3b.
As can be seen in the figure, however, we find no clear correlation between them ($R=-0.29$ for independent 13 regions with significant detection of Br$\alpha$, Br$\beta$, $\mathrm{H_2O}$ ice, and $\mathrm{CO_2}$ ice), which suggests that ice-forming clouds do not significantly contribute to obscuring the observed hydrogen recombination lines; the corresponding HII regions are likely to be located in the foreground of the ice-forming clouds.
Then the column densities of the ices can be underestimated if the HII regions contribute to the continuum emission via free-free emission.
Assuming that 50\% of the continuum emission comes from the foreground, we confirm that the column densities of the both $\mathrm{H_2O}$ and $\mathrm{CO_2}$ ices in each spectrum increase by factors of 2.1--2.4, but without significantly changing their relative abundances.


We examine the difference in the spatial distribution between $\mathrm{H_2O}$ and $\mathrm{CO_2}$ ices.
Figure~3c shows the $\mathrm{CO_2/H_2O}$ ice abundance ratios plotted against the PAH 3.3 $\micron$ intensities.
In the figure, we find that there is a tight correlation between them ($R=+0.88$, $N=14$).
As shown in Fig.~2, the spatial distribution of the PAHs has a sharp peak and monotonically decreases from the galactic center.
Thus the result suggests that the spatial distribution of the ices is clearly different between $\mathrm{H_2O}$ and $\mathrm{CO_2}$ ices, and the abundance of $\mathrm{CO_2}$ ice relative to $\mathrm{H_2O}$ ice decreases significantly with the galactocentric distance.
Here it is notable that the $\mathrm{CO_2/H_2O}$ ratios are strongly correlated with the PAH intensities, whereas the total ice column densities are not (Fig.~3a), indicating that $\mathrm{CO_2}$ ice is efficiently formed in the galactic center.


What causes the change in the $\mathrm{CO_2/H_2O}$ ice abundance ratios?
Figure~3d shows the ice abundance ratios plotted against the PAH 3.3 $\micron$/Br$\alpha$ flux ratios.
In the figure, we find a negative correlation between the $\mathrm{CO_2/H_2O}$ ice abundance ratios and the PAH 3.3 $\micron$/Br$\alpha$ ratios ($R=-0.78$, $N=14$); the $\mathrm{CO_2/H_2O}$ ice abundance ratios are high in HII-dominated regions, while they are small in PDR-dominated regions.
Since the minimum photon energy to ionize hydrogen ($>$ 13.6 eV) is higher than that to excite PAHs ($<$ 13.6 eV), contribution from massive stars to the UV radiation field are likely to reflect the PAH 3.3 $\micron$/Br$\alpha$ ratios.
Then the result implies that massive stars are important to enhance the $\mathrm{CO_2/H_2O}$ ice abundance ratios.
The galactic center of M~82 is indeed characterized by intense starburst (Westmoquette et al. 2007; Keto et al. 1999), which may cause the high $\mathrm{CO_2/H_2O}$ ice abundance ratios in the galactic center.


As for the formation processes of $\mathrm{CO_2}$ ice, energetic processes are favorable in M~82 rather than non-energetic ones, because $\mathrm{CO_2}$ ice formation is enhanced by the intense radiation environment (Fig.~3d).
However we should note that interstellar UV photons cannot penetrate into dense molecular clouds, where ices are formed.
Even in such a situation, UV photons can be induced by cosmic-rays penetrating deep inside clouds through their interactions with molecular hydrogen (Prasad \& Tarafdar 1983).
Although the acceleration process of cosmic-rays has not yet been fully understood, it is likely that many supernova remnants due to the starburst activities increase the cosmic-ray energy density in the galactic center.
In fact, in our Galaxy, significant increases in cosmic-ray ionization rate are reported for several molecular clouds near the Galactic center and some supernova remnants (Goto et al. 2008; Indriolo et al. 2010; Indriolo \& McCall 2012; Geballe \& Oka 2010), which support our idea.
In Fig.~3d, the data points for NGC~253 are also plotted against the PAH 3.3 $\micron$/Br$\alpha$ ratios.
From the figure, we find that they are distributed in a narrower range of the PAH 3.3 $\micron$/Br$\alpha$ ratios than for M~82, which can explain why no significant variations in the $\mathrm{CO_2/H_2O}$ ice abundance ratios were observed for NGC~253 (Yamagishi et al. 2011).
Finally, for comparison with more intense starburst galaxies, we re-analyzed the $AKARI$ NIR spectra of 5 (ultra)luminous galaxies (NGC~34, CGCG~436-030, ESO~507-G070, Arp~193, and Arp~220; Imanishi et al. 2010; Lee et al. 2012).
The resultant $\mathrm{CO_2/H_2O}$ ice abundance and PAH 3.3 $\micron$/Br$\alpha$ ratios are in ranges of 0.08--0.17 and 6.7--19.2, respectively, which follow the trend in Fig.~3d, but in similar ranges to those for M~82.
This may suggest similarity in the hardness of the UV radiation field between M~82 and these galaxies as a whole.


\section{Conclusions}

With $AKARI$, we have obtained the spatially-resolved near-infrared (2.5 -- 5.0 $\micron$) spectra for the nearby starburst galaxy M~82.
We clearly detect spectral features due to $\mathrm{H_2O}$ and $\mathrm{CO_2}$ ices, the PAH 3.3 $\micron$ feature, Br$\alpha$, and Br$\beta$. 
Based on the column density maps of $\mathrm{H_2O}$ and $\mathrm{CO_2}$ ices created from the spectral fitting, we show a significant difference in the spatial distribution between $\mathrm{H_2O}$ and $\mathrm{CO_2}$ ices on a galactic scale; $\mathrm{CO_2}$ ice is more concentrated near the galactic center.
We find a strong negative correlation between the $\mathrm{CO_2/H_2O}$ ice abundance ratios and the PAH 3.3 $\micron$/Br$\alpha$ ratios, indicating that the $\mathrm{CO_2/H_2O}$ ice abundance ratios are high in HII-dominated regions, while they are small in PDR-dominated regions.
The result suggests that massive stars are important to enhance the $\mathrm{CO_2/H_2O}$ ice abundance ratios.
Increase of cosmic-ray induced UV photons due to the intensive starburst activities may cause the high $\mathrm{CO_2/H_2O}$ ice abundance ratios in the galactic center.

\acknowledgments

We would like to thank all the members of the $AKARI$ project.
This work is based on observations with $AKARI$, a JAXA project with the participation of ESA.
We also express many thanks to the anonymous referee for the useful comments.
This work is supported by Grants-in-Aid for Japan Society for the Promotion of Science Fellows No. 23005457 and Grant-in-Aid for Scientific Research No. 22340043.

\begin{table}
\begin{center}
\caption{Summary of the observations and sub-slit apertures used in the present study} 
\centering 
\begin{tabular}{ccccc} 
\tableline\tableline 
Obs. date   & Obs. ID & slit & sub-aperture size & region \\ 
\tableline 
2008 Oct. 21,22 & 3390001.1, 2 & Nh & $3\arcsec\times4\farcs4$ & A \\
2008 Oct. 23 & 3390002.1 & Nh & $3\arcsec\times7\farcs3$ & C\\
2008 Oct. 22 & 3390003.1 & Nh & $3\arcsec\times7\farcs3$ & B\\
\tableline 
\end{tabular}
\end{center}
\end{table}

\begin{figure}
\epsscale{0.48}
\plotone{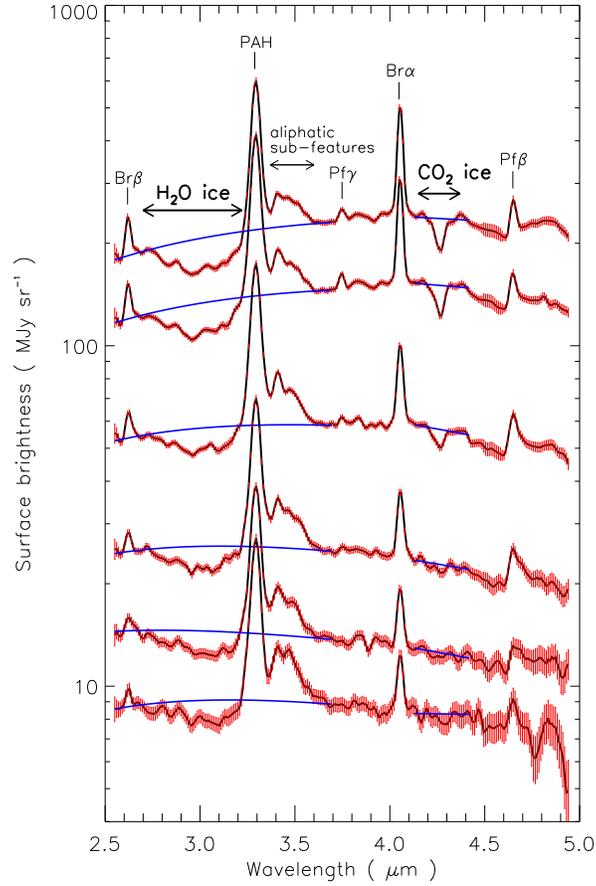}
\figcaption{Variations of the NIR spectra in the center of M~82 (region A, defined in Fig.~2a). The spectra are  obtained from the sub-apertures placed at an interval of 4$\farcs$4 along the slit from the galactic center (top) to the south-east direction (bottom). The blue curves at 2.5--3.7 $\micron$ and 4.1--4.4 $\micron$ in each spectrum indicate the best-fit continuum levels used to evaluate the absorption features due to $\mathrm{H_2O}$ and $\mathrm{CO_2}$ ices, respectively. }
\end{figure}

\begin{figure}
\epsscale{0.48}
\plotone{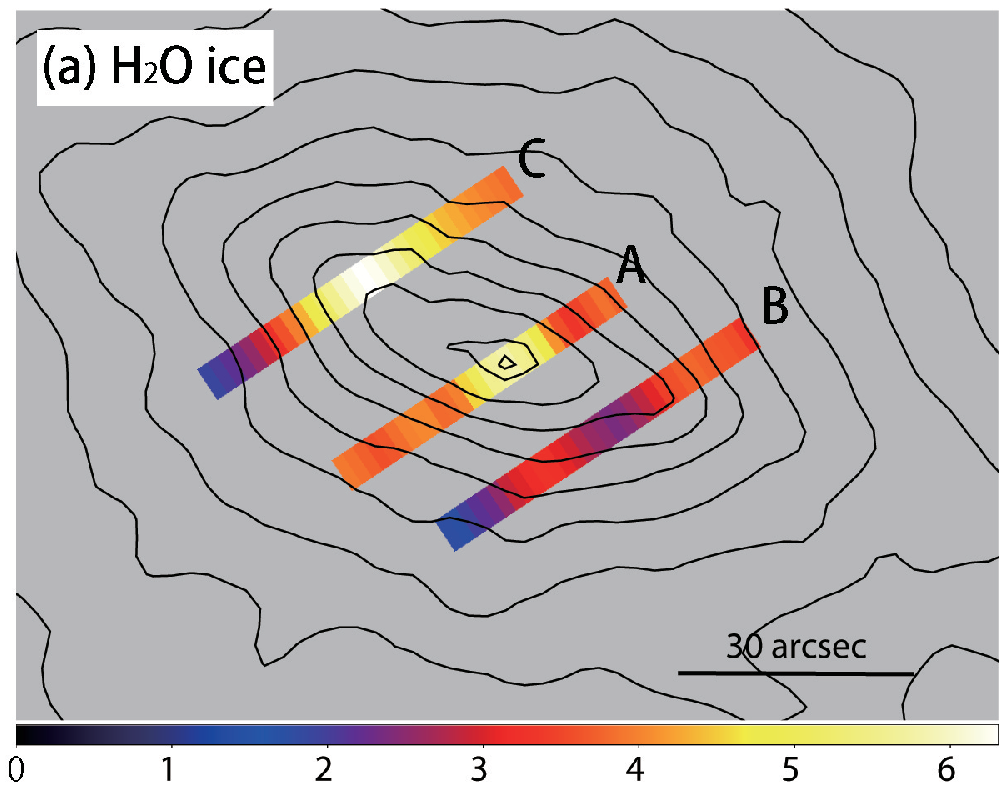}
\plotone{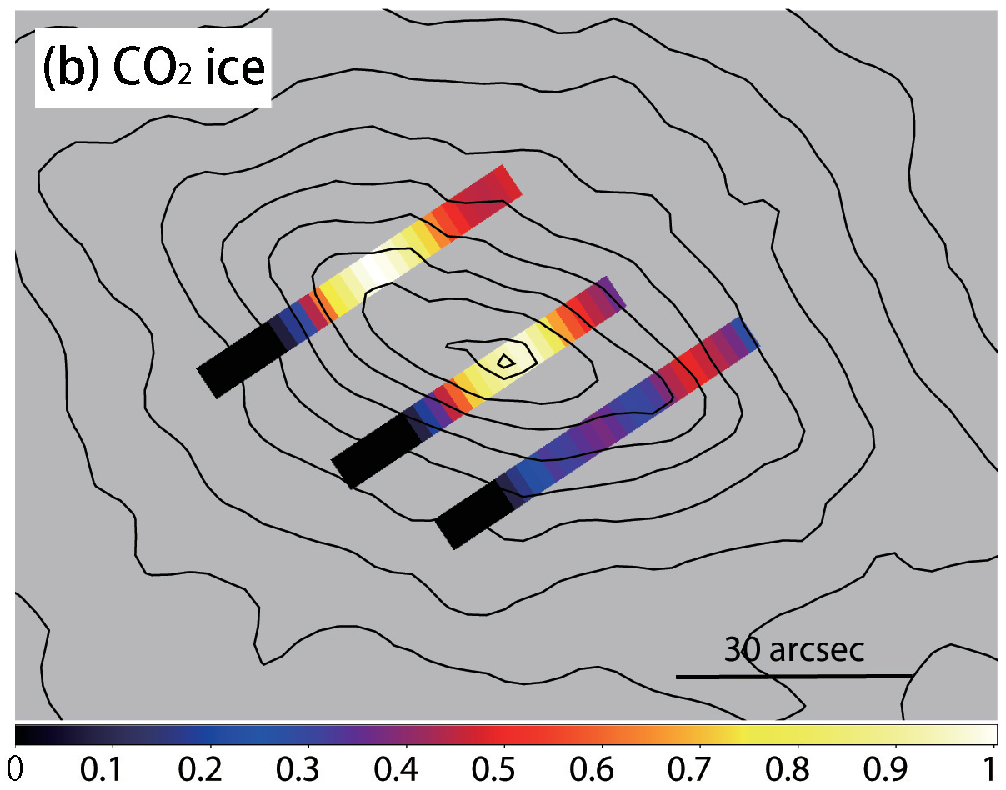}
\plotone{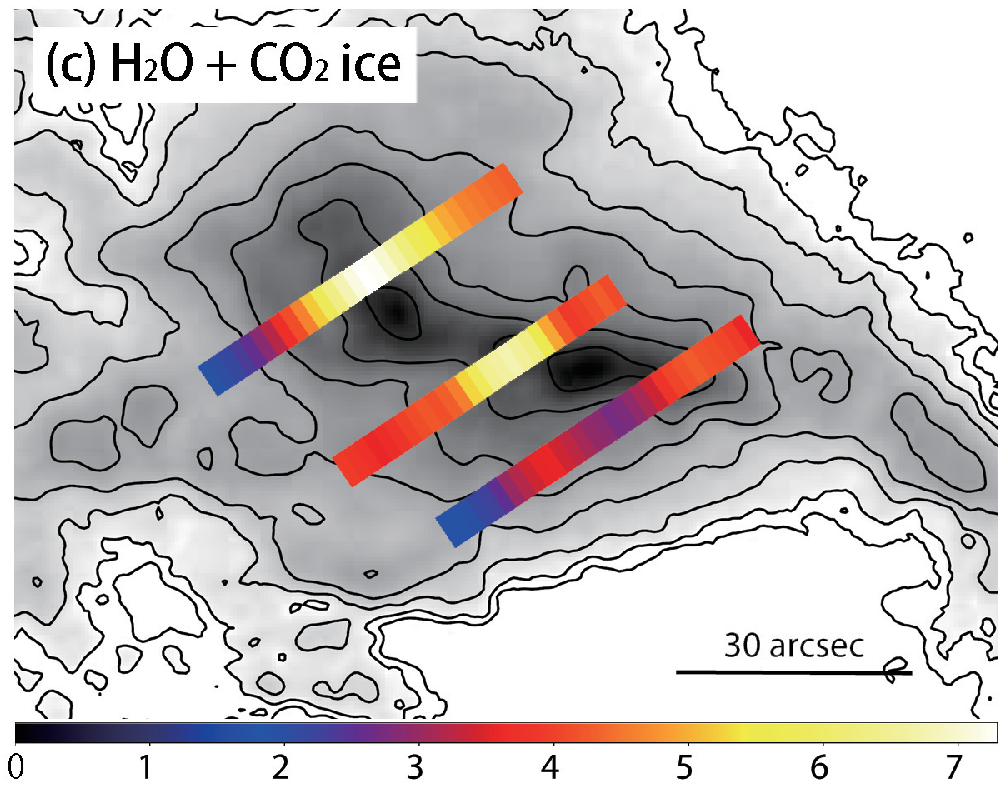}
\figcaption{Maps of the column densities of (a) $\mathrm{H_2O}$ ice, (b) $\mathrm{CO_2}$ ice, and (c) $\mathrm{H_2O}$+$\mathrm{CO_2}$ ice for M~82. The color levels are given in units of $10^{17}$ $\mathrm{cm^{-2}}$. North is up and east is to the left. The contours in Figs.~2a and 2b are taken from the $AKARI$ 7 $\micron$ image (Kaneda et al. 2010). The background image in Fig.~2c is the $^{12}\mathrm{CO}$ ($J$=1-0) map from Walter et al. (2002).}
\end{figure}

\begin{figure}
\epsscale{0.48}
\plotone{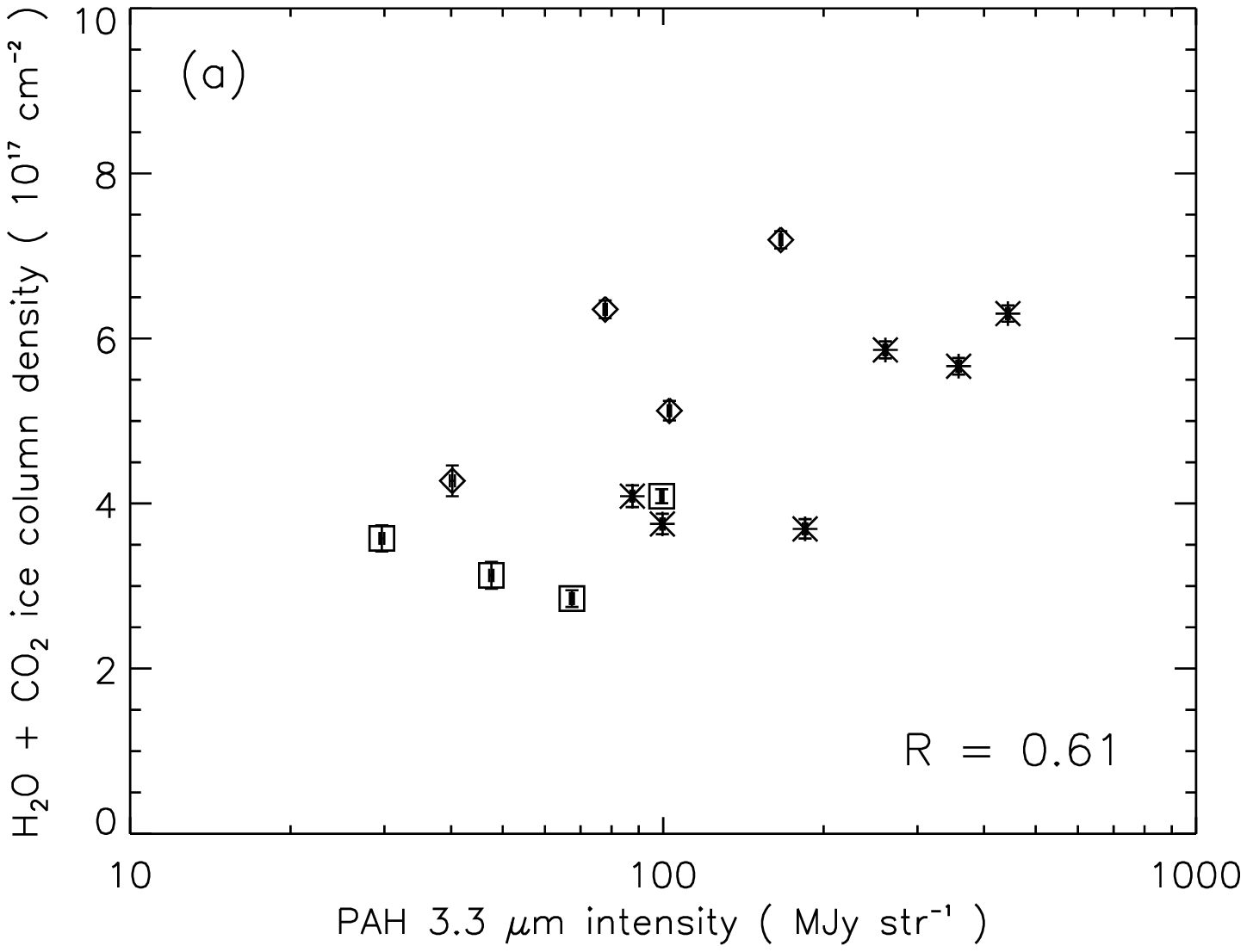}
\plotone{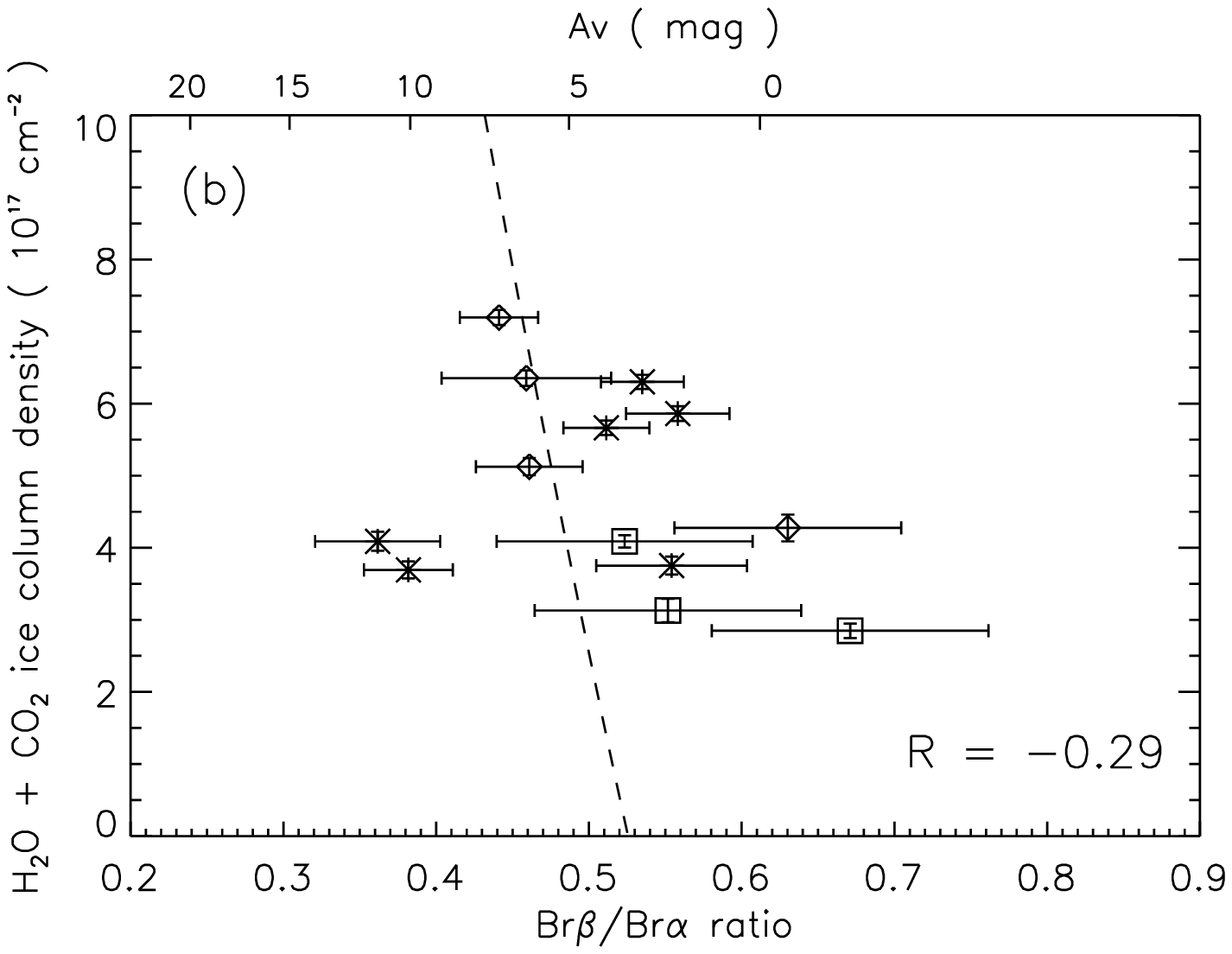}
\plotone{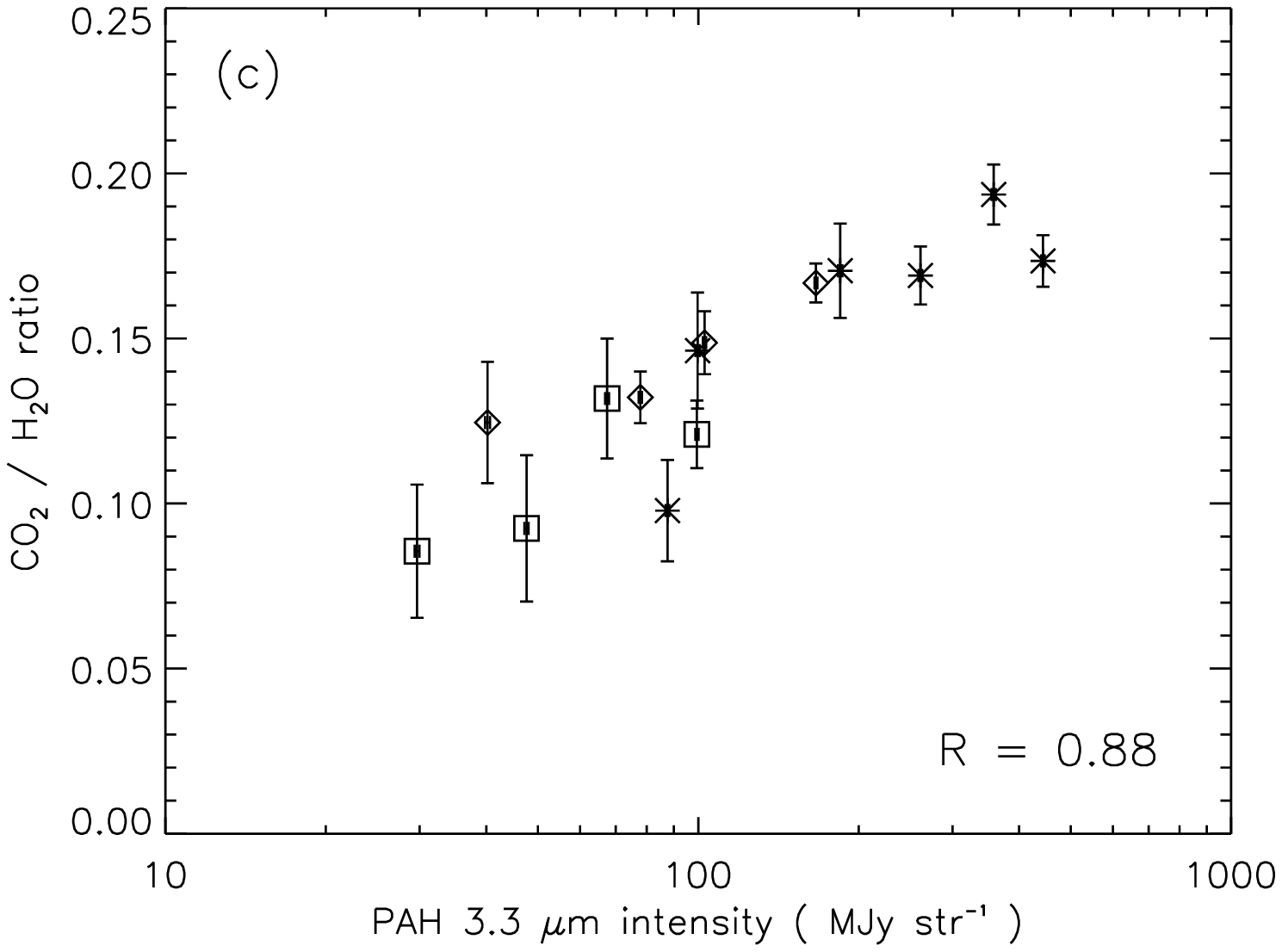}
\plotone{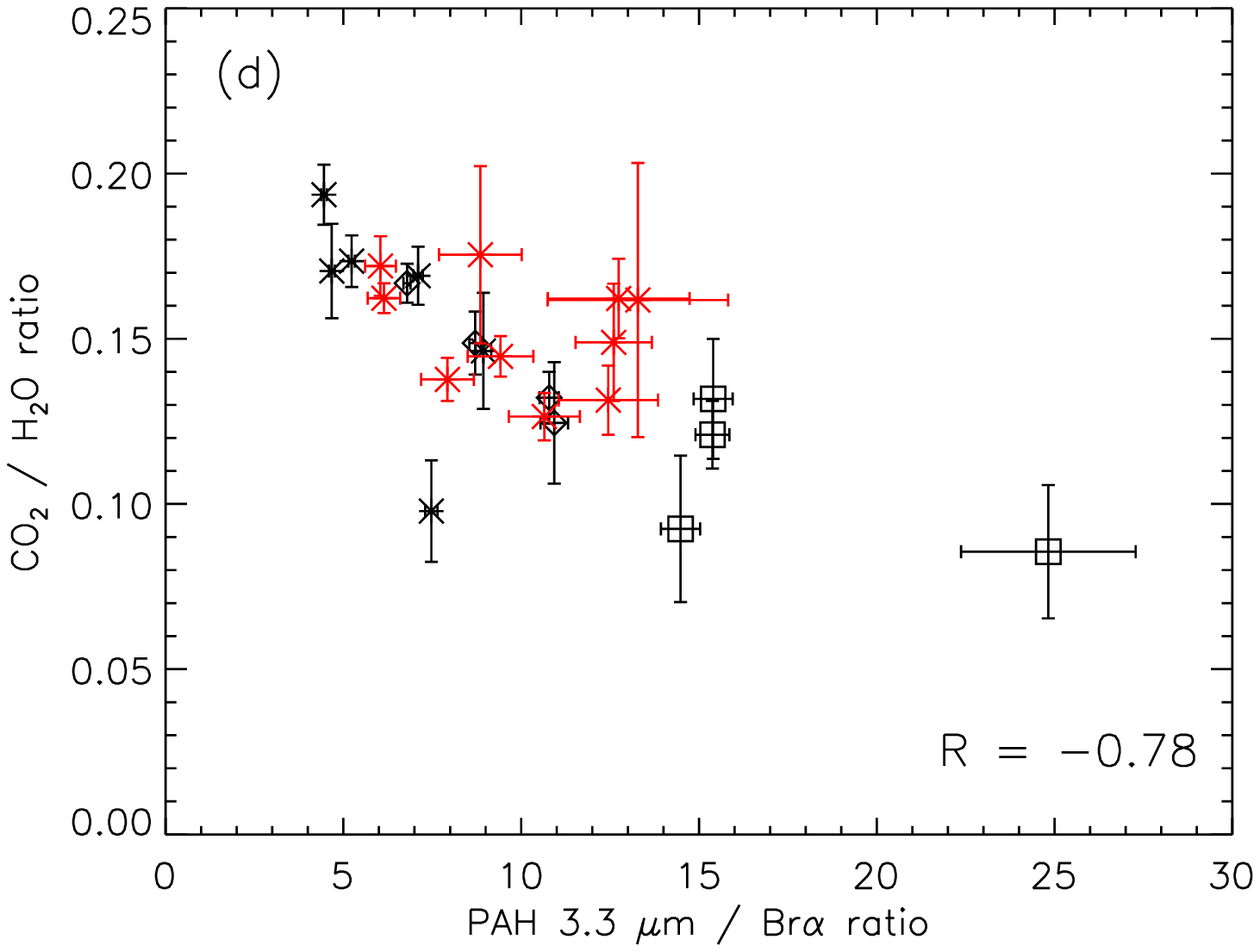}
\figcaption{$\mathrm{H_2O}$+$\mathrm{CO_2}$ ice column densities plotted against (a) the PAH 3.3 $\micron$ feature intensities and (b) Br$\beta$/Br$\alpha$ ratios. $\mathrm{CO_2/H_2O}$ ratios plotted against (c) the PAH 3.3 $\micron$ features intensities, and (d) the PAH 3.3 $\micron$/Br$\alpha$ ratios. All the errors are estimated from the spectral fitting. 
The black asterisks, squares, and diamonds represent the data points in regions A, B, and C, respectively. The correlation coefficient for each plot is shown at the bottom right. The dashed curve in Fig.~3b represents the relation between $A_{\mathrm{V}}$ and $\mathrm{H_2O}$+$\mathrm{CO_2}$ ice column densities in our Galaxy (Whittet et al. 1988; 2007). The red asterisks in Fig.~3d represent the data points for NGC~253 (Yamagishi et al. 2011). }
\end{figure}

\end{document}